\documentstyle[prl,aps,preprint,eqsecnum]{revtex}
\newcommand{\ds}{\displaystyle}
\begin{document}

\title{Wave Propagation in Gravitational Systems: \\
Late Time Behavior}

\author{E.S.C. Ching${}^{(1)}$, P.T. Leung${}^{(1)}$,
W.M. Suen${}^{(2)}$, and K. Young${}^{(1)}$}

\address{${}^{(1)}$Department of Physics,
The Chinese University of Hong Kong, Hong Kong}
\address{${}^{(2)}$Department of Physics, Washington University,
St Louis, MO 63130, U S A}

\date{March 22, 1995}

\maketitle

\begin{abstract}

It is well-known that the dominant late time behavior of waves
propagating on a Schwarzschild spacetime is a power-law tail;
tails for other spacetimes have also been studied. This paper
presents a systematic treatment of
the tail phenomenon for a broad class of models via a Green's
function formalism and establishes the following.
(i)  The tail is governed by a cut of the frequency Green's
function $\tilde G(\omega)$ along the $-$~Im~$\omega$ axis,
generalizing the Schwarzschild result.
(ii) The $\omega$ dependence of the cut is determined by the
asymptotic but not the local structure of space.
In particular it is independent of the presence of a horizon,
and has the same form for the case of a star as well.
(iii) Depending on the spatial asymptotics, the late
time decay is not necessarily a power law in time. The Schwarzschild
case with a power-law tail is exceptional among the class of
the potentials having a logarithmic spatial dependence.
(iv) Both the amplitude and the time dependence of the
tail for a broad class of models are obtained analytically.
(v) The analytical results are in perfect agreement with
numerical calculations.

\end{abstract}

\draft

\pacs{PACS numbers: 04.30.-w}

\section{Introduction}

In this paper we are concerned with the systematics of
the propagation of linearized waves on curved background,
in particular, the late time behavior \cite{prl2}.
The problem is of interest for understanding wave phenomena in
relativity\cite{1,2,3,4,5}, and possibly in other areas of physics.
The propagation of linearized gravitational,
electromagnetic and scalar waves is often modeled by the
Klein-Gordon (KG) equation with an effective potential
\begin{equation}
D \phi(x,t) \equiv \left[ {\partial_{t}^{2}} - {\partial_{x}^{2}}
+V(x) \right] \phi (x,t) = 0  ,
\end{equation}
which includes the Zerilli and Regge-Wheeler equations as special
examples\cite{regge}.
The effective potential $V(x)$ describes
the scattering of $\phi$ by the background curvature. We are
here interested in the evolution of $\phi$ in the presence of
this scattering.

On flat space ($V=0$), waves travel along the light cone:
$\phi (x,t) = \phi_+(x-t) + \phi_-(x+t)$, but the situation is much
more interesting in the presence of a nontrivial background. Fig.~1
shows the result of a typical numerical experiment evolving~(1.1)
on $0 \leq x < \infty$ (so that $x$ may be thought of as a radial
variable, see below), with the boundary condition $\phi (x=0,t)=0$,
and the condition of outgoing waves for $x \rightarrow \infty$. An
initial gaussian $\phi$ centered at $y_o$ with initial $\dot{\phi} = 0$
is allowed to evolve, and is observed at $x_1>y_o$.
The figure shows $\phi(x_1,t)$ versus $t$.  Our interest in this paper
is the behavior at large $t$, but it is useful to first place the
time dependence in an overall context.

The generic time dependence of waves observed at a fixed spatial
location $x_1$ consists of three components, as illustrated in Fig.~1.
(i)  The prompt contribution (peaks $A$ and $C$ in Fig.~1a)
depends strongly on the initial conditions. The first peak $A$ is due
to the initial pulse propagating from $y_o$ directly to the right,
while the later peak $C$ is due to the initial pulse propagating from
$y_o$ to the left, being reflected and inverted at the origin, and then
propagating to the observation point $x_1$.  (ii)  Each of these pulses
is followed by a trailing signal (respectively $B$ and $D$) caused by
the potential $V$ (in the sense that $B$ and $D$ vanish when $V=0$).
The latter pulse $D$ has traversed regions of larger $V$, and as a result
is both more pronounced in amplitude and also more compressed in time.
On a longer time scale (Fig.~1b), it is seen that scattering by the
potential (e.g. peak $D$) merges into a train of quasinormal ringing,
decaying in an exponential manner. (iii)~Finally,
there is a tail, which decays approximately as a power in $t$ (a straight
line in this log-log plot), and it dominates at late times.

The prompt contribution is most intuitive, being the obvious counterpart
of the light-cone propagation in the $V=0$ case, and is in any event
expected in the semi-classical limit.  The ringing is due to a
superposition of quasinormal modes [which are solutions to (1.1) with
the time dependence $e^{-i\omega t}$ for some complex $\omega$].
These are particularly interesting in that the time dependence (both
of the quasinormal modes and of the trailing signals $B$ and $D$) observed
far away carries information about the structure of the intervening space.
In the mathematical sense, this is because the quasinormal mode frequencies
$\omega$ are determined by $V(x)$.  In a more physical sense, it is useful
to consider the analogy of electromagnetic waves emitted by a source in
an optical cavity; the cavity, in providing a nontrivial environment, is
analogous to the presence of a potential $V(x)$ in (1.1).  [In fact, a
one-dimensional scalar model of electromagnetic waves would be described
by a wave equation with a nontrivial dielectric constant distribution
$n^2(x)$ \cite{sculli}; such a wave equation would be very similar
to (1.1) and in fact can be transformed into it \cite{waveeq}.]  For a
broad band source in a laser cavity of length $L$, the spectrum of
electromagnetic waves observed far away would have frequency components
$\omega \approx  j \pi c / L$, where $j$ is an integer.  In this sense,
these quasinormal modes carry information about the cavity.  In much the
same way, the ringing component of the solution in (1.1), if observed,
would carry significant information about the background curvature of
the intervening space.  For this reason, there have been extensive numerical
studies of various models.  In addition, one particularly intriguing
possibility is that the sum of quasinormal modes may be complete in a
certain sense, as the normal modes of a conservative system are complete.
If this were the case, one would have a discrete spectral representation
of the dynamics, with obvious conceptual and computational advantages.
The completeness has been demonstrated in the analogous case of the wave
equation \cite{complete}, and its generalization to the KG equation (1.1)
will be reported elsewhere\cite{QN ringing}.

This paper develops a general treatment for the late time behavior of a
field satisfying (1.1) with a broad class of time-independent potentials
$V(x)$ and the outgoing wave boundary condition. Such a system could
describe linearized waves evolving in stationary spacetimes with or
without horizons, and in vacuum or nonvacuum spacetimes.
For example, for a general static spherically symmetric spacetime,
\begin{equation}
ds^{2} = g_{tt}(r)dt^{2} + g_{rr}(r)dr^{2} + r^{2}(d\theta ^{2}
+ \sin ^{2}\theta  d \varphi ^{2}) \ \ \ ,
\end{equation}
a Klein-Gordon scalar field $\Phi$ can be expressed as
\begin{equation}
\Phi = \sum_{lm} {1\over r} \phi _{lm} (x,t) Y_{lm}(\theta ,\varphi ) .
\end{equation}
The evolution of $\phi_{lm}(x,t)$ is given by (1.1)
with the effective potential
\begin{equation}
V(x) = - g_{tt} {l(l + 1)\over r^{2}} -{1\over 2r} {g_{tt}\over g_{rr}}
\left[{\partial_r g_{tt} \over g_{tt}} -
{\partial_r  g_{rr} \over g_{rr} } \right] \ \ \ ,
\end{equation}
where $x$ is related to the circumferential radius $r$ by
\begin{equation}
x = \int \sqrt{-{g_{rr}\over g_{tt}}} dr
\end{equation}
and extends over a half-line if the metric is non-singular, but extends
over the full line when there is a horizon.

The special case of linearized perturbations of a black hole background
has received much attention \cite{3,4,5,regge,8,9}.
In the simplest case of the Schwarzschild spacetime
\begin{equation}
ds^{2} = -\left(1 - {2M\over r}\right) dt^{2} +
\left(1 -{2M\over r}\right)^{-1} dr^{2}
 + r^{2}(d\theta ^{2} + \sin ^{2}\theta  d \varphi ^{2}) \ \ ,
\end{equation}
the transformation (1.5) leads to
\begin{equation}
x = r + 2M \log \left({r\over 2M} -1 \right).
\end{equation}
For a Klein-Gordon scalar field, $V$ is defined by (1.4) and
takes the explicit form
\begin{equation}
V(x) = \left( 1-{2M\over r} \right)
\left[{\l(\l + 1)\over r^{2}} + (1 -S^{2}) {2M\over r^{3}} \right] \ \ ,
\end{equation}
with $S = 0$.  Electromagnetic and gravitational waves also obey (1.1)
and (1.8), but with $S=1,2$ respectively \cite{regge}. The boundary condition
is outgoing waves at $r = 2M$ ($x \rightarrow - \infty$, into the black
hole) and as $r \to \infty $ ($x \rightarrow \infty$).  With the effective
potential (1.8), it is well known that the tail decays as $t^{-(2\l+3)}$
independent of the spin $S$ of the field\cite{3}. For Reissner-Nordstr\"om
black holes, the evolution equation can again be put in form of (1.1), but
with an effective potential slightly different from (1.8). The tail
is again found to decay as $t^{-(2\l+3)}$\cite{4}. Although a similar
tail should exist for Kerr black holes, we are not aware of
the corresponding analysis in that case.  Most of these previous analyses
(except Ref. 3) are specific to a particular form of the potential.

Several recent developments make a more general study of the late time tail
phenomenon of gravitational systems particularly interesting. As
studies of the Schwarzschild spacetime show that the tail comes from
scattering at large radius\cite{3}, it has been suggested\cite{4} that
a power-law tail would develop even when there is no horizon in the
background, implying that such tails should be present in perturbations
of stars, or after the collapse of a massless field which does not
result in black hole formation. In Ref. 6, the late time behavior of
scalar waves evolving in their own gravitational field or in gravitational
fields generated by other scalar field sources was studied numerically,
and power-law tails have been reported (though with exponents different
from the Schwarzchild case). These interesting results call for a systematic
analysis of the tail phenomenon in (i) nonvacuum, (ii) time-dependent,
and (iii) nonlinear spacetimes, for which the present work will lay a
useful foundation.

In Section II, we formulate the problem in terms of the Green's
function, leading to a precise definition of the prompt, quasinormal mode
and tail components in terms of contributions from different parts of
the complex frequency plane.  In Section III, numerical experiments are
presented.  The analytic treatment is then given in Section IV (potentials
decreasing faster than $x^{-2}$), Section V (inverse square potentials)
and Section VI (composite potentials consisting of a centrifugal barrier
plus a subsidiary term).  The last case, though technically most complicated,
is also the most interesting physically, since it includes, for example,
the potential in (1.8). Section VII contains a discussion and the conclusion.

The highlights of this paper are as follows.
Firstly, the analytic results agree perfectly with the
numerical experiments, including both the time
dependence and the magnitude; this agreement indicates that the
phenomenon of the late time tail has now been quite fully understood
for a large class of models.
Secondly, our work extends, places in context, and provides insight
into the known results for the Schwarzschild case.
For the Schwarzschild spacetime, the late time tail has
previously been related to either reflection from the
asymptotic region of the potential \cite{3}, or to
the Green's function having a branch cut along the negative
imaginary axis in the complex frequency plane \cite{9}.
In this paper, we demonstrate that the cut is a general feature,
if $V(x)$ asymptotically $(x \to \infty)$ tends to zero less rapidly
than an exponential, except for the particular case of the
``centrifugal barrier" $V(x) = \l(\l + 1)/x^{2}, \: l = 1, 2...$.
Furthermore, we determine the strength of the cut in terms of
the asymptotic form of the potential for a
broad class of potentials.  These results
then relate the two different explanations of the late time tail
in the literature.

Some specific late time behavior we discovered in this paper are of
particular interest. Potentials going as a
centrifugal barrier of angular momentum $l$ ($l$ is an integer)
plus~$\sim x^{- \alpha} (\log x)^{\beta}$ ($\beta = 0,1$)
are of interest, because these include the Schwarzschild case
($\alpha = 3, \beta = 1$).  For such potentials, the
generic late time behavior is $\sim t^{-(2l + \alpha)} (\log t)^\beta$.
Thus, for $\beta = 1$, the late time tail is not a simple power law, but
exhibits an additional $\log t$ factor. However, when $\alpha$ is an
odd integer less than $2l + 3$, the generic leading behavior vanishes.
For $\beta = 0$, the next leading term is expected to be
$t^{-(2l + 2 \alpha-2)}$, while for $\beta = 1$, the
next leading term is $t^{-(2l+\alpha)}$ without a $\log t$ factor.
Interestingly, the well-known Schwarzschild spacetime belongs to such
exceptions.

As a collorary, we establish that the form of the late time tail is
independent of the potential at finite $x$.  In particular, whether
the background metric describes a star (so that $x$ is itself the
radial variable $ 0 \le x < \infty$), or a black hole [so that $x$ is
the variable defined in (1.5), $ - \infty < x < \infty$], the late time
tail is the same as long as the asymptotic potential
$ V (x \rightarrow \infty) $
is the same, thus confirming the suggestion of Ref. 5.

As a by-product, we point out that numerical calculations are sometimes
plagued by a ``ghost" potential arising out of spatial discretization.
The effects of this ``ghost" potential may in some cases mask
the genuine tail behavior --- which is often numerically
very small (see Fig.~1b).  This problem could have led to errors, or
at least ambiguities, in some numerical studies, in particular numerical
studies of tail phenomena.  This subtlety with numerical calculations is
discussed in Appendix A.

\section {Formalism}
The time evolution of a field $\phi (x,t)$ described by (1.1) with
given initial data can be written as\cite{Morse}
\begin{equation}
\phi (x,t) = \int  dy \ G(x,y;t) \ \partial_t \phi(y,0)
+ \int  dy \ \partial_t G(x,y;t) \ \phi (y,0)
\end{equation}
for $t > 0,$ where the retarded Green's function $G$ is defined by
\begin{equation}
DG(x,y;t) = \delta (t) \delta (x- y) \ ,
\end{equation}
and the initial condition $G(x,y;t) = 0$ for $t < 0$.
The Fourier transform
\begin{equation}
\tilde{G}(x,y;\omega ) =\int^{\infty }_{0} dt G(x,y;t) e^{i\omega t}
\end{equation}
satisfies
\begin{equation}
\tilde{D}(\omega ) \tilde{G} \equiv \left[-\omega ^{2} -
\partial_x^{2} +V(x)\right] \tilde{G}(x,y;\omega ) = \delta (x- y)
\end{equation}
and is analytic in the upper half $\omega $ plane.

Waves propagating on curved space and observed by a distant observer
should satisfy the outgoing wave condition.  In terms of this, there
are two classes of systems.

\noindent (a) For non-singular systems, e.g., an oscillating
relativistic star generating gravitational waves, the variable $x$,
as defined by (1.5), is the radial variable of a three-dimensional
problem,  $0 \le x < \infty $. The left boundary condition is taken
to be
\begin{mathletters}
\begin{equation}
\tilde{\phi }(x=0,\omega ) = 0 \ ,
\end{equation}
while the right boundary condition is one of outgoing waves
\begin{equation}
\tilde{\phi }(x,\omega ) \propto  e^{i \omega x} \  \ , \qquad
x \rightarrow  +\infty  \ \ ,
\end{equation}
\end{mathletters}where $\tilde{\phi }$ is the wavefunction in the
frequency domain.

\noindent (b)  For the case of black holes,  $x$ given by (1.5)
is a full line variable, $- \infty  < x < \infty $, where the left
boundary  $x \rightarrow - \infty $ is the event horizon.  We have
outgoing waves at both ends, i.e.,
\begin{equation}
\tilde{\phi }(x,\omega ) \propto  e^{-i \omega x} \  \ ,  \qquad
x \rightarrow  -\infty \ \ ,
\end{equation}
together with (2.5b).

For simplicity, we shall write the formalism in terms of the
half-line problem, and indicate briefly the simple changes necessary
for the full-line case. We shall assume $V \rightarrow 0$ as
$|x| \rightarrow \infty$.

Define two auxiliary functions $f(\omega ,x)$ and $g(\omega ,x)$
as solutions to the homogeneous equation
$\tilde{D}(\omega) f(\omega,x) = \tilde{D}(\omega) g(\omega,x) = 0$,
where $f$ satisfies the left boundary condition and $g$ satisfies
the right boundary condition. To be definite, we
adopt the following normalization conventions:
\begin{equation}
\nonumber
\lim_{x\rightarrow \infty} [e^{-i \omega x} g(\omega ,x)] = 1;
\end{equation}
and for the half-line problem
$f(\omega ,x=0) = 0, f^\prime (\omega ,x=0) = 1$\cite{footnote},
while for the full-line problem
$\lim_{x\rightarrow - \infty } [e^{i \omega x} f(\omega ,x)] = 1$.
Let the Wronskian be
$ W(\omega ) = W(g,f) = g \partial_x f  - f \partial_x g$,
which is independent of $x$. It is then straightforward to
see that
\begin{equation}
\tilde{G}(x,y;\omega )= \cases{
{\ds f(\omega ,x) g(\omega ,y) \over \ds W(\omega )} \ , &  $x < y$ \cr \cr
{\ds f(\omega ,y) g(\omega ,x) \over \ds W(\omega )} \ , &  $y < x$ }
\ \  \ \ .
\end{equation}
The essence of the problem is, in one guise or another, how to
integrate $f$ from the left and $g$ from the right, until they can be
compared at a common value of $x$, at which the Wronskian can then be
evaluated. The above choice of normalization is for convenience only,
since a change of normalization, $f \rightarrow Af$ and
$g \rightarrow Bg$, would result in $W \rightarrow AB \, W$ with
no effect on $\tilde G$.

Consider the inverse transform of (2.3), and for $t > 0$ attempt to
close the contour in $\omega $ by a semicircle of radius $C$ in the
lower half-plane, where eventually one anticipates taking
$C \rightarrow  \infty $.  One is then led to consider the
singularities of $\tilde{G}$ for Im~$ \omega < 0$ (Fig. 2), from
which different contributions to $G$ can be identified as follows.

\noindent {\bf (a) Tail contribution} If the potential $V(x)$ has
a finite range, say $V(x) = 0$ for $x > a$, then one can impose the
condition (2.7) at $x = a^{+}$, and integrate the equation through
a {\it finite} distance to obtain $g(\omega ,x)$
for all $x <$ {\it a}.  Such an
integration over a finite range using an ordinary differential
equation cannot lead to any singularity in $\omega$\cite{12}.
It is not surprising that if $V(x)$ does not have
finite support, but nevertheless vanishes sufficiently rapidly as
$x \rightarrow  +\infty , \ g(\omega ,x)$ is still regular.  The
necessary condition is that $V(x)$ must vanish faster than any
exponential\cite{13}, and this is shown in Section~IV.  However,
if $V(x)$ has a tail that does not decrease rapidly
enough, then $g(\omega ,x)$ will have singularities in $\omega$.
A tail that is exponential at large $x$, $V(x) \sim e^{- \lambda x}$,
would in general lead to singularities in $g(\omega,x)$ for
$-$~Im~$\omega \ge \lambda/2$. In so far as these singularities
stay away from $\omega = 0$, they have no effect on the late time
behavior. However, if $V(x \to \infty) \sim x^{- \alpha} (\log x)^{\beta}$
$(\beta = 0,1)$, then $g(\omega,x)$ will have singularities that lie
on the $-$~Im~$\omega $ axis, and take the form of a branch cut
going all the way to the origin\cite{13}, as in the Schwarzschild
case\cite{9}. This contribution to $G$ will be denoted as $G_{L}$.
It is the purpose of this paper to evaluate $G_{L}$ for a broad
class of $V(x)$.

For the half-line problem, $f(\omega ,x)$ is
integrated from $x = 0$ through a finite
distance, and hence does not have any singularities in $\omega $.
For the full-line problem,
$f(\omega ,x)$ is dealt with in the same manner. In all cases of
physical interest, $V(x)$ vanishes on the left either faster than an
exponential, or precisely as an exponential [e.g. Schwarzschild spacetime
under the transformation (1.5)]. For the former, $f(\omega,x)$ has no
singularities in $\omega$, while for the latter, there will be a
series of poles, but at a finite distance from $\omega = 0$. In both
cases, there is no contribution to the late time behavior.

\noindent {\bf (b)  Quasinormal modes}  Secondly,
the Wronskian $W(\omega )$ may have zeros at $\omega  =
\omega _{j}, j = \pm 1, \pm 2, \ldots $, on the complex $\omega$
plane\cite{14}.  At these frequencies, $f$ and $g$ are linearly
dependent, i.e., one can find a solution that satisfies {\it both}
the left and the right boundary condition.  Such solutions are by
definition quasinormal modes (QNM's).  The collective contribution
of the QNM's is denoted as $G_{Q}(x,y;t)$. In the case of a
Schwarzschild black hole, $G_Q$ has been evaluated semi-analytically
by using the phase-integral method\cite{Andersson}. Since
each Im~$\omega _{j} < 0, G_{Q}$ decays exponentially as
$t \rightarrow  \infty $, and is, therefore, irrelevant
for the late time behavior. In some cases,
$G_{Q}(x,y;t)$ is the sole contribution
to the Green's function; this would of course be important,
since the quasinormal modes would then be complete. This case
will be discussed elsewhere\cite{QN ringing}.

\noindent {\bf (c) Prompt contribution}  Finally there is the
contribution from the large semicircle at
$|\omega | = C, C \rightarrow  \infty $.
Since $| \omega | $ is large, this term contributes to the short
time or prompt response, and the corresponding part of the Green
function will be denoted as $G_{P}(x,y;t)$. This term can be shown
to be zero beyond a certain time; thus it does not affect the late
time behavior of the system.

The three contributions $G_{L}, G_{Q}$ and $G_{P}$ can be
understood heuristically using a space-time diagram of wave
propagation, as illustrated in Fig. 3, in which free
propagation is represented by $45^{\circ}$ lines, and the action
of $V(x)$ is represented by scattering vertices.  The initial
wavefunction is assumed to have compact support, so the propagation
starts at some initial $y$, and one imagines observation at some
{\it x}. Rays 1a and 1b show ``direct" propagation without scattering.
These rays arrive promptly at $x$, and correspond to $G_{P}$.
The ray labelled as 2 suffers repeated scatterings at finite {\it x}.
For ``correct" frequencies, the multiple reflections add coherently,
so that retention of the wave is maximum.  Moreover, the amplitude
decreases in a geometric progression with the number of scatterings,
which is in turn proportional to the time {\it t}.  Thus $\phi $
decreases exponentially in $t$; these contributions correspond
to $G_{Q}$.

Lastly, for a wave from a source point $y$ reaching a distant
observer at $x$, there could be reflections at very large $x'$,
such as indicated by ray 3.
The ray first propagates to a point $x' \gg x$, is scattered by
$V(x')$, and returns to the observation point $x$, arriving at a
time $t \simeq (x'-y)+(x'-x) \simeq 2 x'$.
If the potential $V(x) \sim x^{-\alpha}(\log x)^\beta$
with $\alpha \ge 2$, then the scattering amplitude
goes as $V(x') \simeq V(t/2) \sim t^{- \alpha} (\log t )^\beta$.
Thus this contribution leads to $G_L$.
In Section~V, we shall see that this heuristic argument does not
work for $\alpha= 2$. In particular, a centrifugal barrier of angular
momentum $l$ corresponds to free propagation in three-dimensional space,
and does not contribute to the late time tail.

This heuristic picture requires two modifications, which we shall
demonstrate both numerically and analytically in the following
Sections.
First, for $V(x) = l(l+1)/x^2 + \overline{V}(x)$ as $x \to \infty$,
where $\overline{V}(x) \sim x^{-\alpha} (\log x)^{\beta}$, the
late time tail is due to $\overline{V}(x)$ and turns out to be
$t^{-(2l+ \alpha)} (\log t)^{\beta}$ generically.
The suppression by a factor $t^{-2l}$, at least in the case
$\alpha = 3$, is known for specific black hole
geometries \cite{3,4}.  Second, if $\alpha$ is an odd integer
less than $2l+3$, the leading term in the late time tail
vanishes.  For $\beta=0$, the next leading term is expected
to be $t^{-(2l+2 \alpha -2)}$, while for $\beta=1$,
the next leading term is $t^{-(2l+\alpha)}$ without a
$\log t$ factor.

A minor technical complication should be mentioned at the outset.
We close the contour by a semicircle of radius $C$ (or other
large contour). For any finite $C$, one can define the contributions
from (a) the cut, (b) the poles and (c) the large circle. When $C
\rightarrow \infty$, the sum of these must tend to the finite limit
$G(x,y;t)$; however, there is no {\it a priori} guarantee that each
of the three terms would individually converge to some limit. It can
be shown\cite{QN ringing} that provided a regulation is introduced,
then each term would individually approach a finite limit, and hence
the tail, QNM and prompt contributions would each be well defined.
The regulator could be $e^{-\omega \tau}, \tau \rightarrow 0^+$
(for the part coming from Re $\omega \ge 0$). The need for a regulator
comes from the region $|$Im~$ \omega| = O(\log$ $|$Re $\omega|)$,
$|$Re $ \omega | \rightarrow \infty$, and has no bearing on the
tip of the cut, near the origin, which controls the late time behavior.
Therefore, the regulating factor will not be written out in the rest of
this paper.

To study the late time behavior, it then remains to study the spatial
asymptotics of the potential and the consequent singularities of $g$.
Before carrying out these analyses for various potentials in Sections~IV
to VI, we first explore the late time behavior numerically.

\section{Numerical Calculations}

In this Section, we present numerical investigations
of the late time behavior.
Eq. (1.1) is integrated numerically in time using standard
techniques. Instead of the usual second-order differencing
for the spatial derivative term, which could lead to erroneous results
on account of a subtlety to be discussed in Appendix~A, we have used
a higher-order differencing scheme.

We evolve $\phi(x,t)$ for the half line $x \in [0, \infty)$,
with the boundary conditions $\phi =0$ at $x=0$ and
outgoing waves for $ x \rightarrow \infty$.  The full line case
[$x \in (-\infty, \infty)$, with outgoing wave boundary conditions for
$|x| \rightarrow \infty$] is basically the same. From (2.1), the generic
leading late time behavior is given by the integral $\int dy G(x,y;t)
\partial_t \phi(y,0)$, so we take
$\phi(y,t=0) = 0$ and use gaussian initial data:
$\partial_t \phi(y,t=0) = e^{-(y-y_o)^2/\eta^2}$.

We study a broad class of potentials:
\begin{equation}
V(x) = {\nu(\nu+1) \over x^2} + \overline{V}(x)  \ \ ,
\end{equation}
with
\begin{equation}
\overline{V}(x) = {x_o^{\alpha-2} \over x^{\alpha}}
\left[ \log \left({x \over x_o} \right) \right]^\beta \ \ ,
\qquad \qquad x > x_c \ \ ,
\end{equation}
for some $\alpha > 2$, and some $x_o$ and $x_c$,
which includes (i) power-law potentials ($\beta=0$)
and (ii) logarithmic potentials ($\beta=1$).
We study both integral and non-integral $\nu$.
The logarithmic potential is interesting as it
includes the Schwarzschild metric as
a special case.
For $x \le x_c$, we round off the potential by taking
\begin{equation}
V(x) = {2 \nu(\nu+1) \over (x^2 + x_c^2)} +
{2 x_o^{\alpha-2} \over (x^\alpha + x_c^\alpha)}
\ \left[ \log \left( { x+x_c \over 2 x_o} \right) \right]^\beta \ \ ,
\qquad \qquad x \le x_c \ \ .
\end{equation}
We have checked that the form of the late time behavior does not
depend on the round-off.
For all the numerical calculations shown below,
we take $x_c = 100$, $y_o=500$ and $\eta=50$.

We first consider power-law potentials $\overline{V}$ with $\nu=0$ and
study the dependence of the late time tails on the parameters
$\alpha$ and $x_o$.
Fig. 4 shows $\log | \phi(x_1,t) |$ versus $\log t$ at a fixed point
$x_1 > x_c$ for different values of
$\alpha$ and $x_o$. The solid lines are numerical results while the
dashed lines are analytical results, which will be derived in the
following Sections. For all cases studied, the tails are of the form
$t^{-\mu}$ with $\mu \simeq \alpha$. For a given $\alpha$,
the magnitude of the tails is larger when $x_o$ is larger.

Next, we look at the case when $\overline{V}(x)$ is zero and investigate
how the late time tail depends on $\nu$. It turns out that the result
depends significantly on whether $\nu$ is an integer.
For non-integral $\nu$, the tail is again of the form $t^{-\mu}$
but with $\mu \simeq 2+2\nu$. On the other hand, when $\nu$ is an integer,
we see only quasinormal ringing and no power-law late time tail.
This qualitative difference in the tail is observed even when $\nu$
is altered by a small value, and this is demonstrated in Fig. 5, which
shows $\log|\phi (x_1,t)|$ versus $\log  t$ for $\nu = 1$ and $1.01$.
This result suggests that the magnitude of the tail is probably
proportional to $\sin \nu \pi$, and therefore vanishes when $\nu$ is
an integer.

Then, we study power-law and logarithmic potentials with
nonzero integral values of $\nu$, i.e., $\nu = l = 1, 2, \ldots$.
Two interesting results are found. First, logarithmic potentials
often lead to logarithmic tails of the form $t^{-\mu} \log t$,
with $\mu \simeq 2l + \alpha$. Thus, the late time tail is not
necessarily an inverse power of $t$. To exhibit the logarithmic factor,
we plot $ |\phi| t ^{2l + \alpha}$ versus $\log t$ for several logarithmic
potentials with $l=1$ in Fig. 6. The existence of a $\log t$ factor is
indicated clearly by the sloping straight lines at late times [see below
for case~(iii), which is an exception]. However, when we vary the
parameter $\alpha$ continuously from $2.9$ to $3.1$, the behavior of the
late time tail changes discontinuously. This is seen from case (iii) in
Fig.~6. Case~(iii) is asymptotically flat, showing that there is
no $\log t$ factor, and that the tail becomes a simple power law
$\sim t^{-(2l + \alpha)}$. Such discontinuous jumps are observed
whenever $\alpha$ goes through an odd integer less than $2l +3$.
Interestingly, the Schwarzchild metric is such an exception, and the
late time tail decays as $t^{-(2l + 3)}$, with no $\log t$ factor,
as is well known.

Discontinuous jumps are also observed for the power-law
potentials at these values of $\alpha$, and the change in behavior
is more pronounced. The generic late time behavior for the
power-law potentials is $t^{-\mu}$ with $\mu \simeq 2l + \alpha$,
but when $\alpha$ goes through an odd integer less than $2l +3$,
$\mu$ jumps to a larger value $\simeq 2l + 2 \alpha - 2$.
One such discontinuous jump is demonstrated in Fig. 7
with $l = 1$ and $\alpha = 2.9, 3.0$ and $3.1$.

These results and other cases studied but not shown here are summarized in
Table 1. In the following sections, we present an analytic
treatment for the various different potentials. The results so obtained
will be shown to agree very well with the numerical evolutions plotted
here.

\section{Potential decreasing faster than $ x^{-2}$}

The differential equation $\tilde D(\omega) g(\omega ,x) = 0$,
together with the boundary
condition at $x = +\infty $, leads to the exact integral equation
\begin{equation}
g(\omega ,x) = e^{i\omega x} - \int^{\infty }_{x} dx' \,
{\sin \omega (x-x')\over \omega } V(x') g(\omega ,x') \ .
\end{equation}
In Born approximation, the factor $g(\omega ,x')$ in the integrand
may be replaced by
$e^{i\omega x'}$,
so (4.1) can be reduced to
\begin{equation} g(\omega ,x) = e^{i\omega x} - I(\omega ,x)  \ ,
\end{equation}
where
\begin{equation}
I(\omega,x) = \int_x^{\infty} dx' \, { \sin \omega(x-x^\prime) \over
\omega} V(x^\prime) e^{i \omega x^\prime} \ .
\end{equation}
Note that if $V(x)$ is of finite range, then there can be no
singularities in $\omega$.

Suppose the potential decreases like an exponential, i.e.,
$V(x) \propto  e^{-\lambda x}$, then we have
\begin{equation}
I(\omega ,x) \propto  { e^{(- \lambda + i \omega) x }
\over \lambda (\lambda -2i\omega)}  \ ,
\end{equation}
implying a pole at $\omega  = - i\lambda /2$
on the $-$~Im~$\omega $ axis.
Heuristically, if the
potential decreases faster than any exponential, then
$\lambda  \rightarrow  \infty $ and the pole
effectively disappears, resulting in an analytic $g(\omega,x)$.
As a result, there is no tail contribution to the Green's function.
For the exponential potential, $g(\omega,x)$ can actually be solved
exactly without using the
Born approximation\cite{13}. The pole structure and the
outgoing wave condition can then be seen explicitly\cite{remark}.
In any event, since these singularities stay away from $\omega = 0$,
they will not contribute to the late time behavior.

Next, we consider potentials decreasing less rapidly than an
exponential, and in particular $\nu =0$ and a power-law $\overline{V}$,
viz.,
\begin{equation}
V(x) = {K \over x_o^2} \left( {x_o \over x} \right)^{\alpha} \ .
\end{equation}
We take $\alpha > 2$ and the coefficient $K$ will be absorbed into
the scale parameter $x_{o}$.  When $\alpha$ equals an integer $n$,
we have (Appendix B)
\begin{eqnarray}
\nonumber
I(\omega ,x) =  e^{-i \omega x} && \left \{
{(2i \omega x_{o})^{n-2}\over {(n-1)!}} \left[
\gamma  - \sum_{m=1}^{n-2} {1 \over m} + \log (- 2i\omega x) \right]
\right. \\
&& \left. +{1 \over n-1} \left({ x_o \over x} \right)^{n-2}
\sum^{\infty }_{m=0, m \ne n-2}
{(2i\omega x)^{m}\over m! (m+2-n)} \right \} \ ,
\end{eqnarray}
where $\gamma $ is Euler's constant.
The appearance of $e^{-i \omega x}$ in $I$ and thus in $g(\omega,x)$
indicates that a right-propagating wave at
$x = +\infty $, when extrapolated to finite $x$, contains a small
admixture of left-propagating wave.
With reference to ray 3 in Fig. 3, it is clear that this is the term
responsible for the late time behavior.
The logarithm in (4.6) causes a branch cut
along the $-$~Im~$ \omega $ axis (see Fig. 2).
To describe the cut, let
$$
g_{\pm }(-i\sigma ,x) =
\lim_{\epsilon \rightarrow o^{+}} g(-i\sigma \pm \epsilon ,x)
$$
for $\sigma $ real and positive, and
\begin{eqnarray}
\nonumber
\Delta (\sigma ,x) &&= g_{+}(-i\sigma ,x) -  g_{-}(- i\sigma ,x) \\
&&= - \lim_{\epsilon \rightarrow 0^+}
 \left[ I(-i\sigma+\epsilon,x) - I(-i \sigma-\epsilon,x) \right] \ .
\end{eqnarray}
Then from (4.2) and (4.6), the strength of the cut is
\begin{equation}
\Delta (\sigma , x) \simeq 2\pi i {(2\sigma x_{o})^{n-2}
\over {(n-1)!}} e^{-\sigma x} \ \ .
\end{equation}

For non-integral $\alpha$, we have (Appendix B)
\begin{equation}
I(\omega,x) = {-e^{-i \omega x}
 \over (\alpha-1)} \left[ (-2i \omega x_o)^{\alpha-2}
\Gamma(2-\alpha) - \left({x_o \over x} \right)^{\alpha-2}
\sum_{m=0}^{\infty} { (2i \omega x)^m \over m! (m+2-\alpha)} \right] \ .
\end{equation}
In this case, the term $(-2i \omega x_o)^{\alpha-2}$ causes a cut, again
along the $-$~Im~$\omega$ axis. The strength of the cut is
\begin{equation}
\Delta(\sigma,x) \simeq
2\pi i {(2\sigma x_{o})^{\alpha-2}
\over \Gamma(\alpha)} e^{-\sigma x} \ \ ,
\end{equation}
which reduces to (4.8) when $\alpha \to n$.
In this sense, the result for non-integral $\alpha$ is more general.

The Green's function $\tilde{G}(x,y;\omega )$ as given by (2.8)
will likewise have a cut along the $-$~Im~$ \omega $ axis.
The values on either side are given by
\begin{equation}
\tilde{G}_{\pm }(x,y;\omega )
= {f(\omega ,y) g_{\pm }(\omega ,x)\over W(g_{\pm },f)} \ \ ,
\end{equation}
where henceforth we take $y <$ {\it x}.  A little manipulation
gives the discontinuity
\begin{eqnarray}
\nonumber
\Delta \tilde{G}(x,y;\sigma ) & \equiv &
\tilde{G}_{+}(x,y;\omega ) -
\tilde{G}_{-}(x,y;\omega )\mid _{\omega =-i\sigma }\\
&=& {W(g_{-},g_{+})\over W(g_{+},f) W(g_{-},f)} f(\omega ,x)
f(\omega ,y)\mid _{\omega =-i\sigma } \ .
\end{eqnarray}
{}From (4.2) and (4.9)
\begin{equation}
W(g_{- },g_{+}) |_{\omega = -i \sigma} \simeq -{ 2\pi i\over x_{o}}
{(2\sigma x_{o})^{\alpha-1} \over \Gamma(\alpha)} \ .
\end{equation}
Since $W(g_{-},g_{+})$ is independent of $x$, in order to derive
(4.13) it is only necessary to know $g_{\pm}$ at {\it one} value
of $x$. For a potential decreasing faster than $x^{-2}$, the Born
approximation is valid for very large $x$, and we have used the
resultant (4.9) only in that region. Nowhere else do we rely on
the Born approximation.  Therefore, these results are {\it exact},
and do not suffer from any inaccuracies arising from the use of the
first Born approximation. In other words, the $t \to \infty$ behavior
relates only to scattering at $x' \to \infty$, for which the first
Born approximation is accurate, provided $\alpha > 2$.
It remains to evaluate $W(g_{\pm },f)$.  To be specific, consider
the half-line problem and evaluate the Wronskian at $x = 0,$
\begin{equation}
W(g_{\pm },f) = g_{\pm }(\omega ,x=0) \ .
\end{equation}
Substituting (4.13) and (4.14) in (4.12), we have
\begin{equation}
\Delta \tilde{G}(x,y;\sigma)
\simeq - { 2 \pi i\over x_{o}}
{(2\sigma x_{o})^{\alpha-1} \over \Gamma(\alpha) }
{f(- i\sigma ,x) f(- i\sigma ,y) \over
g_+(-i \sigma,0) g_-(-i \sigma,0)} \ .
\end{equation}

The contribution arising from the cut along the
$-$~Im~$ \omega $ axis is
\begin{equation}
G_{L}(x,y;t) = -i \int^{\infty }_{0} {d\sigma \over 2\pi }
\Delta \tilde{G}(x,y;-i\sigma ) e^{-\sigma t} \ .
\end{equation}
The $t \rightarrow  \infty $ behavior is
controlled by the $\sigma  \rightarrow  0$
behavior of (4.15).
A little arithmetic then leads to
\begin{equation}
G_L(x,y; t) \simeq
- {2^{\alpha-1} x_{o}^{\alpha-2} \over t^\alpha}
{f(0,x) f(0,y) \over g(0,0)^2} \ .
\end{equation}

Let the initial conditions be $\phi (y,0) = \phi _{o}(y), \
\partial_t \phi (y,0) = \phi _{1}(y)$,
and for simplicity we assume $\phi _{o}, \phi _{1}$ to have
compact support, or at least to vanish at infinity faster than
any exponential. Then the late time behavior is
\begin{equation}
\phi (x,t) \simeq  - { 2^{\alpha-1} x_{o}^{\alpha-2} \over t^\alpha}
\left[ \int^{\infty }_{0} dy f(0,y)
\phi_{1}(y) \right] {f(0,x) \over g(0,0)^2}  \ .
\end{equation}
The term due to $\phi _{o}$ is proportional to
$\partial_t G$; hence it goes as $t^{-(\alpha+1)}$,
and becomes subdominant when $t \rightarrow \infty$.

To compare the theoretical result (4.18) with the numerical
evolutions in Section~III, we obtain the constant $g(0,0)$ and the spatial
function $f(0,x)$ by integrating the time-independent
equation $\tilde D(\omega \to 0) f= \tilde D(\omega \to 0) g = 0$.
The two results agree very well when $\log t$ is large (see Fig. 4).
In Fig. 8, we plot $\phi$ versus $x$
at a time $t_L$ when the $t^{-\alpha}$ behavior is observed, showing
excellent agreement also in the $x$-dependence of $\phi$ at late times.

\section{Inverse square law potential}
In this Section we consider inverse-square-law potentials
($\overline{V}=0$):
\begin{equation}
V(x) = {\nu (\nu  + 1)\over x^{2}} \ \ \ , \ \ \  x > x_c \ .
\end{equation}
\noindent We assume the potential is repulsive and take $\nu  > 0,$
and define $\rho  \equiv  \nu  + 1/2$. For $V \propto x^{-2}$, the Born
approximation is not accurate at {\it any} $x$ (unless the coefficient
of the potential is small), so the derivation in Section IV is not
valid. However, in this case, an explicit solution can be written down.
The solution satisfying the outgoing wave boundary condition as
$x \rightarrow  +\infty $ is exactly
\begin{equation}
g(\omega ,x) = e^{i(\rho +{1\over 2}){\pi \over 2}}
\sqrt {{\pi \omega x\over 2}} H^{(1)}_{\rho }(\omega x) \ .
\end{equation}
Since $H^{(1)}_{\rho }(z) = J_{\rho }(z)$ +  $i N_{\rho }(z)$ and
\begin{mathletters}
\begin{equation}
J_{\rho }(z) \sim {1\over \Gamma (\rho +1)}
({z\over 2})^{\rho } + \ldots  \ \ \ ,
\end{equation}
\begin{equation}
N_{\rho }(z) \sim - {\Gamma (\rho )\over \pi }
({z\over 2})^{-\rho } + \ldots  \ \ \ ,
\end{equation}
\end{mathletters}the origin is a branch point for non-integral $\nu$.
The symmetry in the present problem dictates that the cut should be
placed along the $-$~Im~$ \omega $ axis,
and some algebra leads to
\begin{mathletters}
\begin{eqnarray}
g_{+}(- i\sigma ,x)& =&\sqrt{ {i\pi \sigma x \over 2}}
e^{i\nu \pi /2} [ 2 \cos \rho \pi  H^{(1)}_{\rho }(i\sigma x)
+ e^{-i\rho \pi } H^{(2)}_{\rho }(i\sigma x) ] \ ,
\\
g_{-}(- i\sigma ,x)& =&\sqrt{i\pi \sigma x\over 2} e^{i\nu \pi /2}
[e^{-i\rho \pi } H^{(2)}_{\rho } (i\sigma x)] \ ,
\end{eqnarray}
\end{mathletters}where the functions $H^{(1,2)}_{\rho }(i\sigma x)$
have no discontinuity for $\sigma  > 0.$  Then
\begin{equation}
W(g_{-},g_{+})\mid _{\omega =- i\sigma } =
- 4i\sigma  \sin \nu \pi \ ,
\end{equation}
\noindent which vanishes for integral $\nu $.
This is readily understood since $V(x) = l(l+1)/x^2$
represents a pure centrifugal barrier, which corresponds to free
propagation in three-dimensional space and does not cause any scattering.
Note that for $\nu  \rightarrow  0,$ the potential is weak and
the Born approximation derived in Section IV should be valid.
It can be readily checked that the $\nu  \rightarrow  0$ limit of
(5.5) indeed agrees with the $\alpha \rightarrow  2$
limit of (4.13).  [For $\alpha = 2,$ it is no longer possible to
absorb the magnitude $K$ of the potential into the definition of
$x_{o}$, so (4.13) should be multiplied by
$K = \nu (\nu  + 1) \simeq  \nu $ in this case.]
This agreement may be regarded as a check on the formalism in
Section IV. Thus the dependence $W(g_{- },g_{+})
\propto  \sigma ^{n-1}$ remains valid for $n = 2,$ but the
amplitude vanishes.

For $\omega  = - i\sigma  \rightarrow  0,$
\begin{mathletters}
\begin{equation}
g_{+}(- i\sigma ,x) \sim -{1\over \sqrt{\pi}}
e^{i(3 \rho /2+1/4)\pi}
\Gamma (\rho ) \left({i\sigma x\over 2}\right)^{ -\nu } \ ,
\end{equation}
\begin{equation}
g_{-}(- i\sigma ,x) \sim {1\over \sqrt{\pi}}
e^{i(- \rho /2+1/4)\pi}
\Gamma (\rho ) \left({i\sigma x\over 2}\right)^{- \nu } \ ,
\end{equation}
\end{mathletters}for $x > x_c$. Since $g_{\pm } \propto  x^{-\nu }$,
\begin{equation}
W(g_{\pm },f) =
\left[f'(x_c)+ {\nu \over x_c} f(x_c) \right] g_{\pm}(x_c) \ \ ,
\end{equation}
and we have evaluated the Wronskian at $x=x_c$.
As discussed in Section~II, $f(\omega,x)$ is analytic in $\omega$,
which implies $f(\omega,x) = \sum_{n=0}^{\infty} f_n(x) \omega^n$.
{}From the Wronskian, it can be seen that $f_0(x) \ne 0$. Thus,
\begin{equation}
\lim_{\omega = -i \sigma \rightarrow 0} W(g_{+},f) W(g_-,f) =
- \left[ f_0^{'}(x_c) + {\nu \over x_c} f_0(x_c) \right]^2
{\Gamma(\rho)^2 \over \pi}
\left( {\sigma a \over 2} \right)^{-2 \nu} \ \ \ .
\end{equation}

Using (4.12), we have
\begin{equation}
\Delta \tilde{G}(x,y;\sigma ) \sim
{4i\pi \sin \nu \pi \over \left[ f_0^{'}(x_c) +
(\nu / x_c) f_0(x_c) \right]^2 }
\left({x_c\over 2}\right)^{2\nu }
{\sigma ^{2\nu +1} \over \Gamma (\nu +{1 \over 2})^{2}} f_0(x) f_0(y) \ \ ,
\end{equation}
and the late time behavior is
\begin{equation}
G_L(x,y;t)
\simeq  {2 \sin \nu \pi \over \left[ f_0^{'}(x_c) + (\nu / x_c)
f_0(x_c) \right]^2} {\Gamma (2\nu +2)\over
\Gamma (\nu +{1\over 2})^{2}}
\left({x_c\over 2}\right)^{2\nu } t^{-(2+2\nu)} f_0(x) f_0(y) \ \ .
\end{equation}
Two features are worthy of remark.  First, whereas the lowest order
Born approximation in Section~IV would have predicted a
time-dependence of $t^{-2}$ for a potential $V \sim x^{-2}$, the power
is in fact $t^{-(2+2\nu )}$
for non-integral $\nu$.  For $\nu  \ll 1,$ the two agree,
whereas for $\nu  = O(1)$, the difference arises from higher-order
Born approximation. Secondly, the power-law dependence in $t$ vanishes
for integral $\nu $ and the late time behavior is then dominated by the
exponentially decaying QNM's.

The late time behavior is
\begin{equation}
\phi(x,t)
\sim {2 \sin (\nu \pi)
\over \left[ f_0^{'}(x_c) + \nu f_0(x_c) / x_c
\right]^2} {\Gamma (2\nu +2)\over \Gamma (\nu +{1\over 2})^{2}}
\left({x_c\over 2}\right)^{2\nu } t^{-(2+2 \nu)} f_0(x)
\int_0^\infty dy f_0(y) \phi_1(y)  \ \ .
\end{equation}
This analytical result for $\phi(x,t)$ is evaluated
and plotted as the dashed line for $\nu = 1.01$ in Fig. 5.
The dashed line is indistinguishable from the solid
line (obtained from numerical evolution) when
$\log t \ge 8$. On the other hand, for $\nu=1$,
$\phi$ decreases much faster than
$t^{-(2+2\nu )}$, in fact faster than any power law,
as derived above.

\section{ Composite potential}

We now consider potentials of the following form:
\begin{equation} V(x) = {l(l+1) \over x^2} +
\overline{V}(x), \qquad  x> x_c \ , \qquad \qquad l = 1, 2, \ldots
\end{equation}
with $\overline{V}(x)$ given in (3.2).
We first take $\alpha > 2$ to be non-integer, and
results for integral $\alpha$ will be obtained by taking the
limit $\alpha \rightarrow n$, where $n$ is a integer.
For $l=0$ and $\beta = 0$, $V(x)$
reduces to the case discussed in Section IV.

We see from Section~V that if $V$ is treated as a perturbation,
the lowest order Born approximation no longer gives correct
results. Rather, we should consider $\overline{V}(x)$ as the
perturbation.  The solution for the inverse square potential,
$l(l+1)/x^2$, is given exactly by
\begin{equation}
g^{(0)}(\omega,x) \equiv i^{l+1}
(\omega x) h_l^{(1)}(\omega x) \ \ ,
\end{equation}
where $h_l^{(1)}$ is
the spherical Hankel function of the first kind. Applying the
Born approximation now gives, in analogy to (4.2),
\begin{equation}
g(\omega,x) \simeq  g^{(0)}(\omega,x) - I(\omega,x) \ \ ,
\end{equation}
where
\begin{equation}
I(\omega,x) \equiv
\int_x^\infty dx' M(x,x';\omega) \overline{V}(x')
g^{(0)}(\omega,x') \ \ ,
\end{equation}
with the zeroth-order Green's function  being
\begin{equation} M(x,x';\omega) = {i \omega \over 2}  x x'
\left[ h^{(1)}_l(\omega x') h_l^{(2)}(\omega x)
- h_l^{(1)}(\omega x) h_l^{(2)}(\omega x') \right] \ \ ,
\end{equation}
and $h_l^{(2)}$ is the spherical Hankel function of the
second kind.

First consider a power-law $\overline{V}$.
After some algebra, we get
\begin{equation}
I={-i^{l} \omega x \over 2} \left[
 I_1(\omega,x)  h_l^{(1)} (\omega x) -
 I_2(\omega,x) h_l^{(2)}(\omega x) \right] \ \ ,
\end{equation}
where
\begin{eqnarray}
&& I_1(\omega,x) = (\omega x_o)^{\alpha-2}
\int_{\omega x}^{\infty} du \
{h_l^{(1)}(u) h_l^{(2)}(u)  \over u^{\alpha -2} } \ \ , \\
&& I_2(\omega,x) = (\omega x_o)^{\alpha -2} \int_{\omega x}^{\infty}
du \ { h_l^{(1)}(u)^2 \over u^{\alpha-2}} \ \ .
\end{eqnarray}
While $I_1$ is straightforwardly evaluated to be
\begin{equation}
I_1(\omega,x) = 2^{-2l} (\omega x)^{-2l-1} \left( {x_o \over x }
\right)^{\alpha -2} \sum_{m=0}^{l} a_m(l,\alpha) (\omega x)^{2m} \ ,
\end{equation}
$I_2$ is somewhat more difficult to obtain.
We find
\begin{eqnarray}
\nonumber
I_2(\omega,x)= && (-1)^{l+1} 2i C(l,\alpha)
{ \Gamma(2-\alpha) \over (\alpha -1)}
(-2 i \omega x_o)^{\alpha-2} \\
&& + (-1)^{l}
(\omega x)^{-2l-1} \left( {x_o \over x} \right)^{\alpha-2}
\sum_{m=0}^{\infty} b_m(l,\alpha) (\omega x)^m  \ \ ,
\end{eqnarray}
where
\begin{equation}
C(l,\alpha) =
\prod_{j=0}^{l-1} { \alpha-2j-3 \over \alpha+1+2j} \ \ ,
\qquad l = 1,2, \ldots \qquad ,
\end{equation}
and $C(l,\alpha) = 1$ for $l=0$.
[See Appendix C for the derivation of (6.9) and (6.10) and
the definitions of $a_m$ and $b_m$.]
The term $(-2 i \omega x_o)^{\alpha-2}$ causes a cut in
$g(\omega,x)$, which lies along the negative imaginary axis,
with a strength
\begin{equation}
\Delta(\sigma,x) \simeq \pi (-i)^{l+1} (2 \sigma x_o)^{\alpha-2}
{C(l,\alpha) \over \Gamma(\alpha)}
(2 \sigma x) h_l^{(2)}(-i \sigma x) \ \ .
\end{equation}
Thus the factor $W(g_-,g_+)$ is given by
\begin{equation}
W(g_-,g_+)|_{\omega = -i \sigma} \simeq W(g^{(0)},\Delta)
= - 4 \pi i \sigma (2 \sigma x_o)^{\alpha-2}
{C(l,\alpha) \over \Gamma(\alpha) } \ \ \ .
\end{equation}
In the case of $\alpha$ being an integer $n \ge 3$,
we get
\begin{equation}
W(g_-,g_+) = -4 \pi \sigma (2 \sigma x_o)^{n-2}
{C(l,n) \over (n-1)!} \ \ .
\end{equation}

Approximate $W(g_{\pm},f)$ by $W(g^{(0)},f)$, and we have
$$\lim_{\sigma \rightarrow 0}
W(g_{+},f)W(g_-,f) = \sigma^{-2l} g_o^2 \ \ ,$$
where $g_o \equiv \lim_{\omega \to 0} [(i \omega)^l W(g^0,f)]$
is a non-vanishing constant and reduces to $g(0,0)$ for $l=0$. Thus,
\begin{equation}
\Delta \tilde{G}(x,y;-i \sigma) \simeq - \pi i
2^{\alpha} x_o^{\alpha-2}
\sigma^{2l + \alpha -1}
{C(l,\alpha) \over  \Gamma(\alpha)}
{f(-i \sigma,x) f(-i \sigma,y) \over g_o^2} \ \ ,
\end{equation}
and as a result,
\begin{eqnarray}
G_L(x,y;t)
&&\simeq  - {f(0,x) f(0,y) \over g_o^2}
C(l,\alpha) 2^{\alpha} x_o^{\alpha-2}
{\Gamma(2l + \alpha) \over \Gamma(\alpha)}
t^{-(2l+\alpha)} \cr
&& \equiv - {f(0,x) f(0,y) \over g_o^2}
C(l,\alpha) {F(\alpha) \over t^{2l + \alpha}}
\end{eqnarray}
for both integral and non-integral $\alpha$.

We note the interesting result that
$C(l,\alpha) = 0$ when $\alpha$ equals an odd integer
less than $2l+3$, in which case
the late time tail vanishes in first
Born approximation, and higher order corrections have
to be considered. This implies multiple scatterings from
asymptotically far regions are important in such cases.
We shall not go into the details of the higher order calculations,
but from dimensional analysis, we expect
\begin{equation}
G_L(x,y;t) \sim t^{-(2l + \alpha)}
\left[ A_1 + A_2 \left({x_o \over t} \right)^{\alpha-2} +
A_3 \left( { x_o \over t}\right)^{2(\alpha-2)}  + \cdots \right] \ \ ,
\end{equation}
where $A_i$ 's are  constants  proportional  to
the $i^{\hbox{th}}$-order scattering amplitude.

Generically the late time behavior arising from the first Born
approximation is linear in the potential. The potential
with a logarithmic $\overline{V}$
is obtained by taking $(-\partial/ \partial \alpha)$
on that with a power-law $\overline{V}$. So we immediately obtain
from (6.16) that for logarithmic $\overline{V}$
\begin{equation}
G_L(x,y;t)
= \left\{ C(l,\alpha)F(\alpha) {\log t \over t^{2l+\alpha} } -
{\partial \over \partial \alpha}
\left[ C(l, \alpha) F(\alpha) \right]
{1 \over t^{2l+\alpha}} \right \} {f(0,x) f(0,y) \over g_o^2} \ \ .
\end{equation}
Hence, the leading terms are $t^{-(2l + \alpha)} (c \log t + d)$,
except that $c \propto C(l,\alpha)$ vanishes when
$\alpha$ is an odd integer less than $2l+3$.

Analytical results obtained from (6.16) and (6.18)
with {\it no} adjustable parameters [except in case (ii) in
Fig. 7; see below] are plotted as
dashed lines in Figs. 6 and 7. The agreement between numerical evolutions
and analytical results is perfect.
For case (iii) in Fig. 6, the vanishing of the leading term implies
that the asymptotic slope should be zero (i.e., no log~$t$, but
only a pure power, whose magnitude is determined),
and this indeed agrees with the numerical results,
with no adjustable parameters.
For case (ii) in Fig. 7, the leading term vanishes, and the dashed line
represents the next leading term arising from multiple
scatterings, whose time dependence is determined,
but whose magnitude has been left as an adjustable normalization.
In Fig. 9, we show the dependence of
$\phi$ on $x$ at late times; perfect agreement is again found.

\section{Conclusion}

In this paper we have studied the dynamical evolution of waves on
a curved background that can be described by the Klein-Gordon
equation (1.1) with an effective potential.
This includes linearized scalar, electromagnetic and gravitational
waves on a Schwarzschild background as special cases.  Our work
extends, places in context and provides understanding for the
known results for the Schwarzschild spacetime.

We have given a systematic treatment of waves described by (1.1) using a
Green's function formulation.  The Green's function consists of three
distinct components, $G_L$, $G_Q$, and $G_P$, coming
from the three contributions to the contour integral in
frequency space. Each component leads to a distinct wave phenomenon
in physical spacetime. $G_Q$ leads to quasinormal ringing, $G_P$
leads to prompt responses, and $G_L$
leads to the late time tail phenomenon. Fig. 1 shows clearly these three
contributions, while Fig. 3 provides a heuristic understanding
in terms of wave propagation in a spacetime diagram.

We further focus on the late time tail phenomenon.  Both numerical and
analytic results, which agree perfectly with each other, are presented.
For a broad class of potentials, we have demonstrated
that the tail is related to the Green's function having a branch cut
along the $-$~Im~$\omega$ axis, with the asymptotic late time behavior
controlled by the tip of the cut.
Using a Born analysis, we determine the strength of the cut in terms
of the spatial asymptotics of the potential.  This gives analytic
understanding of both the magnitude and the time dependence of the
late time tail, not just for the Schwarzschild case, but also for a wider
class of models. We have applied our formulation to a few interesting
cases, including the inverse-square-law potential, other power-law
potentials decaying faster than inverse square, potentials containing
a logarithmic term, and composite potentials with a centrifugal barrier.

We find that the well-known late time power law behavior of waves on
a Schwarzschild spacetime turns out to be an exceptional case.  In
general, for potentials going as a centrifugal barrier of angular
momentum $l$ plus $\overline{V}(x) \sim x^{- \alpha} \log x$,
the generic late time
behavior is $\sim t^{-(2l + \alpha)} \log t$, which is not a simple
power law. However, when $\alpha$ is an odd integer less than
$2l + 3$, as in the case of Schwarzschild, the coefficient of the
leading term vanishes, and the next leading term is $t^{-(2l+\alpha)}$
without a $\log t$ factor.  This discontinuous dependence of the time
dependence on $\alpha$ is verified numerically and shown in Fig. 6.

There are various implications, colloraries and interesting future
extensions
discussed throughout the paper. Appendix~A
outlines a subtlety in the numerical calculations of the late time tail
behavior (or more generally speaking the accurate time dependence of
solutions of partial differential equations) in the timelike direction.

\acknowledgements

We thank R. H. Price and H-P. Nollert for discussions.
M.K. Yeung contributed to some aspects of the work in Appendix A.
We acknowledge support from the Croucher Foundation. WMS is also
supported by the CN Yang Visiting Fellowship and
the US NSF (Grant No. 94-04788).

\newpage

\appendix

\section{Ghost potential}

In this Appendix we consider a subtlety in the numerical solution of (1.1),
which we write as
\begin{equation}
\partial ^2_t \phi = L \left[ \phi \right]
\equiv  \partial ^2_x \phi - V \phi \: .
\end{equation}
The numerical solution does {\it not} satisfy (A1), but instead satisfies
a modified equation with $L$ replaced by
\begin{equation}
\widetilde{L} \left[ \phi \right] \equiv D^2_x \phi - V \phi \: ,
\end{equation}
where $D^2_x$ is a finite-difference approximation to $\partial^2_x$, e.g.,
in a second-order scheme,
\begin{eqnarray}
\nonumber
D^2_x \phi(x,t) \: &  =  & \: {\left[  \phi (x+ \Delta x, t) - 2 \phi (x,t)
+ \phi (x - \Delta x, t) \right] \over (\Delta x)^2 } \\
& \simeq & \: \partial^2_x \phi +
\frac{1}{12} (\Delta x)^2 \partial^4_x \phi + \cdots \:  .
\end{eqnarray}
For the class of problems at hand, the solution takes the form
\begin{equation}
\phi (x,t) = t^{-\mu} f(0,x) + \cdots \: ,
\end{equation}
where $\mu$ is some exponent,
$f(\omega, x)$ is the solution regular at the origin, and the
omitted terms are subasymptotic as $t \rightarrow \infty$ at fixed $x$;
see e.g. (6.16).
(The convergence is not uniform in $x$, and moreover, a $\log t$ factor
can be incorporated without changing the argument below.)  To lowest order
in $\Delta x$, this satisfies (A1), and noticing that
$\partial^2 _t \phi$ is down by a factor $t^{-2}$, we see that
$\partial^2_x \phi \simeq V \phi + \cdots$. This allows us
to eliminate all spatial derivatives on $\phi$ higher than second order,
and a little arithmetic leads to
\begin{equation}
\widetilde{L} \left[ \phi \right] \: \simeq \: L \left[ \phi \right]
+ \frac{1}{12} (\Delta x)^2 \left[ (V'' + V^2) \phi
+ 2 V' \phi ' \right] +\cdots \: ,
\end{equation}
where $'$ denotes $\partial_x$.

Two further simplifications are possible.  First, because of (A4),
the $\partial ^2_t \phi$ term is subasymptotic as $t \rightarrow \infty$,
and consequently the error caused by finite differencing in $t$ can be
ignored, at least to order $(\Delta t)^2$.  Secondly, assume that the
potential is dominated at large $x$ by the centrifugal barrier, then
\begin{equation}
f(0,x) \simeq K x^{l+1}
\end{equation}
[The function $f(0,x)$ corresponds to the radial solution to Laplace's
equation, and is regular at the origin.  In general, it contains both
the decreasing and the increasing component.  At large $x$, the latter
dominates, and (A6) then follows. The behavior (A6) has been verified
numerically.]  As a result, $\phi ' \simeq [(l+1)/x] \phi$,
and it is possible to write the extra terms in (A5) purely in terms of
$\phi$ without $\phi '$:
\begin{equation}
\widetilde{L} \left[ \phi \right] \: \simeq \: L \left[ \phi \right]
+ V_{gh}(x) \phi + \cdots \: ,
\end{equation}
where the ghost potential is
\begin{equation}
V_{gh}(x) =  \frac{1}{12} (\Delta x)^2 \left[ V'' + V^2
+ \frac{2(l+1)}{x} V' \right] \propto (\Delta x)^2 x^{-4} \: .
\end{equation}

Thus if $\overline{V}$ vanishes at infinity faster
than $x^{-4}$, say as $x^{-\alpha}$  ($\alpha > 4$),
then instead of seeing the true late time behavior $t^{-(2l+\alpha)}$,
the numerical solution based on second-order spatial differencing would
display a late time behavior $t^{-(2l+4)}$ due to the
ghost potential.  It is easy to be misled by the ghost behavior
unless one has analytic expectations to compare with.

As an even clearer example, consider a pure centrifugal barrier,
say with $l=1$.  The late time behavior should be exponential
(since there should be no cuts along the $-$~Im~$\omega$ axis,
and the leading large $t$ contribution comes from the quasinormal modes).
The top line in Fig. 10 shows the result of a numerical calculation with
second-order spatial differencing; the straight line portion gives a
dependence $t^{-6}$, which is the ghost behavior as explained above.
The lower line in the same figure shows the result when
$\Delta x$ is halved; the $t$ dependence remains unchanged,
but the magnitude is decreased, roughly by a factor of 4, as expected.
The change with $\Delta x$ shows that this tail is a numerical
artifact.

In short, $k$-th order spatial differencing leads to
\begin{equation}
V_{gh} (x) \propto (\Delta x)^k x^{-(2+k)} \ \ ,
\end{equation}
which could mask the effect of $\overline{V}$ if the latter decreases as
$x \rightarrow \infty$ more rapidly.  This problem could have led
to errors, or at least ambiguities, in some numerical experiments.
In all the numerical results presented in this paper, we have taken
care to use spatial differencing of a sufficiently high order that
$V_{gh}$ will not dominate over $\overline{V}$, and have moreover checked
that the $t \rightarrow \infty$ asymptotics claimed are stable against
(a) decreasing $\Delta x$, and (b) increasing the order of spatial
differencing.

\section{}

For a power-law potential of the form (4.5) with $K=1$,
using (4.3), we have
\begin{equation}
I(\omega,x) = \int_x^{\infty} dx' {\sin \omega (x-x') \over \omega}
{x_o^{\alpha-2} \over x^\alpha} e^{-i \omega x'}  \ \ ,
\end{equation}
which is easily shown to be:
\begin{equation}
\nonumber
I(\omega,x) = {(\omega x_o )^{\alpha-2} \over 2i }
\left\{ \left[ \int_{\omega x}^{\infty}
{du \over u^\alpha} \right] e^{i \omega x} -
\left[ \int_{\omega x}^{\infty} { du e^{2iu} \over
u^\alpha} \right] e^{-i \omega x} \right\} \ \ .
\end{equation}
After evaluating the first integral and integrating the
second one by parts, we get
\begin{equation}
I(\omega,x) = - e^{-i \omega x} { (-2i \omega x_o )^{\alpha-2}
\over (\alpha-1) } J(\omega x) \ \ ,
\end{equation}
where
$$ J(\omega x) \equiv \int_{-2i \omega x}^{\infty} du \
{e^{-u} \over u^{\alpha-1}} \ . $$

If $\alpha $ is an integer $n$, we have\cite{Bender}:
\begin{equation}
J(\omega x) = {(-1)^{n-1} \over (n-2)!}
\left[ \gamma - \sum_{m=1}^{n-2} {1 \over m} + \log (-2i \omega x)
\right] + {(-1)^{n-1} \over (2i \omega x)^{n-2} }
\sum_{m=0 \, m \ne n-2}^{\infty} {(2i \omega x)^m \over m! (m+2-n)!} \
,
\end{equation}
where $\gamma$ is the Euler's constant.

On the other hand, if $\alpha$ is non-integer, we have\cite{Bender}
\begin{equation}
J(\omega x) = \Gamma(2-\alpha) + { (-1)^{\alpha-1}
\over (2i \omega x)^{\alpha-2} }
\sum_{m=0}^{\infty} { (2i \omega x)^m \over m! (m+2-\alpha)} \ \ .
\end{equation}

Substituting (B4) and (B5) into (B3) then gives us
(4.6) and (4.9) respectively.

\section{}

The spherical Hankel functions of the first and second kinds
can be written as\cite{table}:
\begin{equation}
h_l^{(1)}(u) = i^{-l-1} u^{-1} e^{iu}
\sum_{m=0}^{l} {(l+m)! \over m! (l-m)!} (-2i u)^{-m} \ \ ,
\end{equation}
and
\begin{equation}
h_l^{(2)}(u) = i^{l+1} u^{-1} e^{-iu}
\sum_{m=0}^{l} {(l+m)! \over m! (l-m)!} (2i u)^{-m} \ \ .
\end{equation}
Using (C1) and (C2), $I_1$ as defined in (6.9) is easily
evaluated to be:
\begin{eqnarray}
\nonumber
 && I_1(\omega,x)   \cr
= && {2^{-2l} \over (\omega x)^{2l+1}}
\left( { x_o \over x} \right)^{\alpha-2}
\sum_{m_1, m_2=0}^{l} { (2i)^{m_1 + m_2}
\left[ (-1)^{m_1} + (-1)^{m_2} \right]
(2l-m_1)! (2l-m_2)! \over
2 (2l+\alpha-1-m_1-m_2) (l-m_1)! (l-m_2)! m_1! m_2!}
(\omega x)^{m_1+m_2}  \ \ .\cr
\end{eqnarray}
Now since $(-1)^{m_1} + (-1)^{m_2}$ vanishes when $m_1$ and $m_2$
are not both odd or both even, we have only even powers of $\omega x$
in the double sum. Hence, we can define the double sum as
\begin{eqnarray}
\nonumber
&& \sum_{m_1=0}^{l} \sum_{m_2=0}^{l} { (2i)^{m_1 + m_2}
\left[ (-1)^{m_1} + (-1)^{m_2} \right]
2 (2l-m_1)! (2l-m_2)!
\over (2l + \alpha -1-m_1 -m_2) (l-m_1)! (l-m_2)! m_1 ! m_2 !}
(\omega x)^{m_1+m_2}  \cr
\equiv && \sum_{m=0}^{l} a_m(l,\alpha) (\omega x)^{2m}  \ \ ,
\end{eqnarray}
and get (6.9).

To evaluate $I_2$, we consider more generally the following integral:
\begin{equation}
{\cal I}(\beta,z) \equiv \int^\infty_z {v(t) \over t^\beta } dt \ \ ,
\end{equation}
where $v(t)$ is an analytic function and
$\lim_{t \rightarrow \infty} v(t) t^n = 0$
for any positive number $n$.  Assume that $v(t=0) \neq 0$, so that
the integrand is singular at $t=0$ for $\beta \ge 0$.
Let the Taylor expansion of $v(t)$ be $\sum^\infty _{n=0} c_n t^n$,
which converges for $t \in [0,\infty)$.

Rewrite ${\cal I}(\beta,z)$ for $\beta \ge 1$ as follows:
\begin{equation}
{\cal I}(\beta,z) =
\int^\infty_z {v(t)-\sum^{N-1}_{n=0} c_nt^n \over t^\beta } dt
+\int^\infty_z {\sum^{N-1}_{n=0} c_nt^{n} \over t^\beta }  \ \ ,
\end{equation}
where $N$ is the largest positive integer less than $\beta$.
It is then straightforward to show that
\begin{equation}
{\cal I}(\beta,z) = \int^\infty_0 {v(t)-\sum^{N-1}_{n=0} c_nt^n
 \over t^\beta} dt
 -\sum^{\infty}_{n=0} {c_nz^{n+1-\beta} \over n+1-\beta }  \ \ ,
\end{equation}
Note that the integrand on RHS is well
behaved at $t=0$. We shall show that this integral is,
in fact, equal to
$$
{\cal I}_A(\beta) \equiv \int^\infty_0 {v(t) \over t^\beta } dt \ \ ,
$$
where the method of analytic continuation has applied to define it for
$\beta \ge 1$.

First, by considering the muliti-valued property of the
function $t^\alpha$ across the branch cut going from $t=0$
to $t=\infty$, one can argue that for noninteger $\beta$
\begin{equation}
{\cal I}_A(\beta) = {e^{\pi \beta i} \over 2i\sin{\beta \pi}}
 \int_C {v(t) \over t^\beta } dt \ \ .
\end{equation}
The integral goes along a contour $C$ enclosing the positive
real $t$-axis in the clockwise direction as shown in Fig. 11. This
representation has the advantage of being analytic in $\beta$
for both positive and negative values of $\beta$
and is readily applicable as an analytic continuation of the
original integral.

Second, by using a similar trick, one can also prove that
\begin{equation}
\int^\infty_0 {v(t)-\sum^{N-1}_{n=0} c_nt^n \over t^\beta } dt =
{e^{\pi \beta i} \over 2i\sin{\beta \pi}}
 \int_C {v(t)-\sum^{N-1}_{n=0} c_nt^n \over t^\beta } dt \ \ .
\end{equation}
Moreover, the integral
$$
\int_C {\sum^{N-1}_{n=0} c_nt^n \over t^\beta } dt
$$
vanishes, which can be seen by deforming the contour $C$ into a
infinitely large circle enclosing the origin, as shown in Fig. 11.
This then completes our proof for the equality
\begin{equation}
{\cal I}_A(\beta) = \int^\infty_0 {v(t)-\sum^{N-1}_{n=0} c_nt^n
\over t^\beta } dt \ \ ,
\end{equation}
and hence
\begin{equation}
{\cal I}(\beta,z) = {\cal I}_A(\beta)
 -\sum^{\infty}_{n=0} {c_nz^{n+1-\beta} \over n+1-\beta }  \ \ .
\end{equation}

{}From (6.8),
\begin{equation}
I_2(\omega,x) = (\omega x_o)^{\alpha-2} J_2(\omega x) \ \ ,
\end{equation}
where
\begin{equation}
J_2(\omega x) \equiv \int_{\omega x}^{\infty} dt
{ \left[ h_l^{(1)}(t) \right]^2 \over t^{\alpha-2} } \ \ .
\end{equation}
We let
\begin{equation}
w(t) = \left[ h_l^{(1)}(t) \right]^2 t^{2l+2} \ \ ,
\end{equation}
so that $w(t)$ is analytic in $t$.
Using (C1), we have
\begin{eqnarray}
w(t) &&= (-1)^{l+1}
\sum_{k_1=0}^{l} \sum_{k_2=0}^{l} \sum_{n=0}^{\infty}
(-1)^{k_1+k_2} (2i)^{n-k_1-k_2}
{(l+k_1)! (l+k_2)! (-1)^{k_1+k_2}
\over k_1! k_2! (l-k_1)! (l-k_2)!}
t^{n+2l-k_1-k_2} \cr
&& \equiv (-1)^{l+1} \sum_{m=0}^{\infty} d_m t^m \ \ .
\end{eqnarray}
Then
\begin{equation}
J_2(\omega x) =
\int_{\omega x}^{\infty} dt {w(t) \over t^{2l+\alpha} } \ \ ,
\end{equation}
which is of the form of (C3) with $v(t)=w(t)$, $\beta = 2l+ \alpha$
and $z = \omega x$.
Hence using (C9), we get
\begin{equation}
J_2(\omega x) = {\cal I}_A + (-1)^l
\sum_{m=0}^{\infty}
{d_m (\omega x)^{m+1-2l-\alpha} \over m+1-2l-\alpha} \ \ ,
\end{equation}
Since ${\cal I}_A$ is the analytic continuation
for the case $\alpha \le -2l$, it can be looked up in
tables\cite{table}:
\begin{equation}
{\cal I}_A = { (-1)^{l+1} (-i)^{\alpha-1} \over 2^\alpha \pi}
{ \Gamma({3-\alpha+2l \over 2}) \Gamma({2-\alpha \over 2})^2
\Gamma({1-\alpha-2l \over 2}) \over \Gamma(2-\alpha)} \ \ ,
\end{equation}
After some algebra, we get
\begin{equation}
{\cal I}_A = (-1)^{l} (-2i)^{\alpha-1} C(l,\alpha)
{\Gamma(2-\alpha) \over (\alpha -1)}
\end{equation}
with $C(l,\alpha)$ defined in (6.11).
Substitute (C15) and (C17) into (C10) and after some simplification,
we obtain (6.10) with
\begin{equation}
b_m \equiv  {d_m \over m+1-2l-\alpha} \ \ .
\end{equation}

\newpage

\centerline{\bf Table 1}

\vskip 25pt
\renewcommand {\arraystretch}{1.76}
\hskip -0.5cm
\begin{tabular}{|c|c|c|} \hline
$V(x), \ \ x \rightarrow \infty$ &
 & $ \phi (t), \ \ t \rightarrow \infty $ \\ \hline
\ $(K/x_o^{2}) (x_o/x)^{\alpha} $ \ & all real $\alpha > 2$ &
$t^{- \alpha}$  \\ \hline
\ $\nu(\nu+1)/x^2 $ & integer $\nu$ \ & $e^{-t/\tau}$ \\ \cline{2-3}
& all non-integer $\nu$ &  $t^{-(2\nu +2)}$ \\ \hline
\ $l(l+1)/x^2 + (K/x_o^{2}) (x_o/x)^{\alpha}$ \ &
\ odd integer $\alpha < 2l +3$ \ &
\ $t^{-\mu}, \ \ \mu > 2l +  \alpha $ \  \\ \cline{2-3}
$l$: integer & all other real $\alpha$ &  $t^{- (2l+ \alpha)} $ \\ \hline
\ $l(l+1)/x^2 + (K/x_o^{2}) (x_o/x)^{\alpha}  \log (x / x_o) $ \
& odd integer $< 2l+3$ & $t^{- (2l + \alpha)} $ \\ \cline{2-3}
$l$: integer & all other real $\alpha$ & $ t^{- ( 2l + \alpha ) } \log t$
\\ \hline
\end{tabular}
\vskip 5pt
\hoffset=0.in
\vskip 30pt
\noindent Table 1. \ Behavior of late time tails for various potentials

\newpage

\centerline{{\bf FIGURE  CAPTIONS} }

\begin{description}

\item[Fig. 1] Generic time dependence of $\phi(x_1,t)$ at a fixed
spatial point $x_1$, evolving
from an initial gaussian $\phi$ and initial $\dot \phi = 0$. The time
evolution is governed by the Klein-Gordon equation (1.1). The potential
in this case is $V(x) \sim l(l+1)/x^2 + (\log x)/x^3$ with $l=1$.
(a) Shorter time scale to show the prompt contributions; (b) longer
time scale to show the late time behavior.

\item[Fig. 2] Singularity structure of $\tilde{G}(x,y;\omega)$ in
the lower-half $\omega$-plane, and the various contributions to the
Green's function $G(x,y;t)$.

\item[Fig. 3] A spacetime diagram illustrating heuristically
the different contributions $G_L$, $G_Q$ and $G_P$ (see text for
definition) to the Green's function.

\item[Fig. 4] $\log |\phi(x_1,t)|$ versus $\log t$ for potentials
with $\nu = 0$ and a power-law $\overline{V}$, which go as
$x_o^{\alpha-2} / x^\alpha$ as $x \to \infty$.
(a) (i) $x_o = 1$, $\alpha =3$; (ii) $x_o = 100$, $\alpha =3$ and
(b) (i) $x_o = 1$, $\alpha = 2.5$; (ii) $x_o = 1$, $\alpha = 5.7$.
Solid lines are numerical results, which agree very well with the
analytical results (dashed lines).

\item[Fig. 5] $\log |\phi(x_1,t)|$ versus $\log t$ for an inverse-square-law
potential $V(x) \sim \nu(\nu+1)/x^2$.
(i) $\nu = 1.01$ and (ii) $\nu = 1$. Solid lines are numerical results.
For $\nu = 1.01$, $\phi$ decreases as
$t^{-\mu}$ with $\mu \simeq 4.03$ which is in excellent agreement with
the analytical result of $t^{-(2+2\nu)}$ (dashed line), derived in
Section~V. For $\nu = 1$, $\phi$ decreases faster
than any power law, again agreeing with the theoretical prediction.
For clarity, case (ii) is shifted downwards by 8.0.

\item[Fig. 6]
$ A | \phi(x_1,t) | t^{2l + \alpha}$ versus $\log t$ for several
potentials $V(x) \sim l(l+1)/x^2 + \log x /x^\alpha$.
(i) $l = 0$, $\alpha = 3$; (ii) $l = 1$, $\alpha = 2.9$;
(iii) $l = 1$, $\alpha = 3 $; (iv) $l = 1$, $\alpha = 3.1$.
For clarity, the data are multiplied by a constant $A$
with (i) $A = 10^{-9}$; (ii) and (iii) $A = 5.6 \times 10^{-10}$;
(iv) $A= 5.86 \times 10^{-10}$. The sloping straight lines observed
at late times in cases (i), (ii) and (iii) indicate the existence of
a $\log t$ factor. Case (iv) is an exception, in which the nearly
zero asymptotic slope shows that the tail is a simple power law.
The numerical evolutions (solid lines) are indistinguishable from
the analytical results (dashed lines) for $\log t > 9.5$.

\item[Fig. 7] $\log |\phi(x_1,t)|$ versus $\log t$ for
various potentials of the form $2/x^2 + 1/x^\alpha$ as $x \to \infty$.
(i) $\alpha = 2.9$; (ii) $\alpha = 3.0$; (iii) $\alpha = 3.1$.
To make the three sets of lines stagger, vertical shifts have been
applied. The generic late time behavior is $t^{- \mu}$ and $\mu
\simeq 2 + \alpha$ except in case (ii) in which
$\mu$ jumps discontinuously to $2 \alpha$. Solid lines are numerical
evolutions while dashed lines are analytical results, derived in
Section~VI. In cases (i) and (iii), the analytical results are completely
specified both in form and magnitude; in case (ii), only the form is
known and the magnitude is fitted to the numerical result.

\item[Fig. 8] $A \phi(x,t_L)$ versus $x$ at a fixed time $t_L$ for the
same potentials plotted in Fig. 4. The numerical evolutions (solid
lines) agree very well with the theoretical estimates (pluses).
(i) $x_o = 1$, $\alpha =3$; (ii) $x_o = 100$, $\alpha = 3$;
(iii) $x_o = 1$, $\alpha = 2.5$; (iv) $x_o = 1$, $\alpha = 5.7$.
The data are multiplied by a constant $A$ for clarity.
(i) $A = 5 \times 10^4$; (ii) $A = 10^6$; (iii) $A = 10^3$;
(iv) $A = 2 \times 10^{15}$.

\item[Fig. 9] $A \phi(x,t_L)$ versus $x$ for various
potentials of the form $2/x^2 + (c_1 \log x + c_2)/x^\alpha$
as $x \to \infty$. Solid lines are numerical results while the
pluses are theoretical estimates.
(i) $c_1 = 1, c_2=0, \alpha =3$; (ii) $c_1 = 0, c_2 = 1, \alpha = 3.3$;
(iii) $c_1 = 0, c_2=1, \alpha = 4$; (iv) $c_1 = 0, c_2=1, \alpha = 5$.
The data are multiplied by a constant $A$ for clarity. (i) $A = 10^8$;
(ii) $A = 10^{10}$; (iii) $A = 10^{12}$; (iv) $A = 10^{15}$.

\item[Fig. 10] Ghost behavior observed in the numerical evolution of
$\phi(x_1,t)$ using a second-order spatial differencing scheme. The
potential used in this case is $V(x) \sim 2/x^2$ as $x \to \infty$ and
the expected late time behavior is an exponential decay as shown in
Fig.~5 (which uses a higher-order spatial differencing scheme) and
Section~V. In the present numerical evolution, however,
a straight line portion indicating a time dependence of $t^{-6}$ is
observed. The magnitude of this ``ghost" tail decreases by a factor of
4 when the spatial interval $\Delta x$ is halved (see Appendix~A),
showing that this tail is a numerical artifact.

\item[Fig. 11] Contour $C$ for evaluating the integral ${\cal I}_A$
defined in (C6).

\end{description}


\begin{references}
\bibitem{waveeq} The wave equation
$ \left[  n^2 (z) {\partial^2 _t}
  -{\partial^2 _z} \right] \psi (z,t) = 0  $
is related to the KG equation (1.1)
by a transformation  $ dx/dz = n(z), \psi  = n^{-1/2} \phi $ ,
with the potential related to $n(z) $ by
$V = (2n^3)^{-1}(d^2 n/dz^2) -(3/4n^4)(dn/dz)^2$ . However, a
continuous $V(x)$ implies a continuous $n(z)$ but not the other
way around.
\bibitem{footnote}
We assume here $f'$ is nonzero at the origin.
Otherwise, if $f \sim x^{l+1}$ as $x \rightarrow 0$, then we take
$\lim_{x \rightarrow 0} x^{-(l+1)} f(\omega,x) = 1$ instead of $f'=1$.
\bibitem{14} It is assumed that these are simple poles;
generalization to higher
order poles, or to cuts, is straightforward. These poles are
always in pairs: $\omega_{-j} = - \omega_{j}^*$.
\bibitem{remark} From Ref. 17 and references therein,
for $V(x) = -V_o e^{-\lambda x}$,
$g$ is exactly solved to be:
$$ g(\omega,x) =
\left({V_o \over \lambda^2} \right)^{i \omega/\lambda}
\Gamma(1-2i \omega/ \lambda)
J_{-2i \omega/\lambda}(2 V_o^{1/2} e^{-\lambda x/2}/\lambda) $$
where $\Gamma$ and $J$ are the
gamma and Bessel functions respectively. The pole at
$\omega = -i \lambda /2$ in the Born approximation [see (4.4)]
corresponds to the first pole of the gamma function when the
argument equals zero.
When $x \rightarrow \infty$, the argument of the Bessel
function becomes small and we have
$\Gamma(1-2i \omega/ \lambda) J_{-2i \omega/ \lambda }
(2 V_o^{1/2} e^{-\lambda x/2}/\lambda)
\simeq (V_o^{1/2} e^{-\lambda x/2}/\lambda)^{-2i \omega/\lambda}
= (V_o / \lambda^2)^{-i \omega/\lambda} e^{i \omega x}$.
Therefore $g(\omega,x) \sim e^{i\omega x}$ when
$x \rightarrow \infty$, satisfying the required outgoing-wave
condition.
\end{references}
\end{document}